# Carrier-envelope phase stable, high-contrast, double-CPA laser system

Aurélie Jullien[1,*], Aurélien Ricci[1,2], Frederik Böhle[1], Jean-Philippe Rousseau[1], Stéphanie Grabielle[3], Nicolas Forget[3], Hermance Jacqmin[1,2], Brigitte Mercier[1], Rodrigo Lopez-Martens[1]

[1]Laboratoire d'Optique Appliquée, École Nationale Supérieure de Techniques Avancées - Paristech, École Polytechnique, CNRS, 91761 Palaiseau Cedex, France
[2]Thales Optronique SA, Laser Solutions Unit, 2 Avenue Gay-Lussac, 78995 Elancourt, France
[3]Fastlite, Les collines de Sophia, 1900 route des cretes, 06560 Valbonne, France
*Corresponding author: aurelie.jullien@ensta-paristech.fr

We present the first carrier-envelope phase stable chirped pulse amplifier (CPA) featuring high temporal contrast for relativistic intensity laser-plasma interactions at 1 kHz repetition rate. The laser is based on a double-CPA architecture including XPW filtering technique and a high-energy grism-based compressor. 8 mJ, 22 fs pulses are produced with $10^{-11}$ temporal contrast at -20 ps and a CEP drift of 240 mrad RMS.
http://dx.doi.org/10.1364/OL.99.099999

Relativistic optics has opened the way to compact plasma-based particle accelerators and XUV and X-rays sources [1–4]. Nowadays, several trends can be recognized in experimental facilities dedicated to intense laser-matter interactions. On the one hand, multi-TW femtosecond lasers with ultra-high intensity can be used to drive compact plasma-based particle accelerators [1, 2]. On the other hand, the ability to generate waveform-controlled few-cycle light pulses has provided brand new tools for investigating ultrafast electronic processes in atoms, molecules, or solids [4]. More recently, few-cycle pulses with relativistic intensity, but without carrier-envelope phase (CEP) control, have been used to study electron bunch acceleration from gas jets or high-harmonic generation from solid targets [5, 6]. At Laboratoire d'Optique Appliquée (LOA), a pioneering experimental platform has been developed (Salle Noire), which merges these two trends, dedicated to the study of laser-plasma interactions in the few-cycle regime at 1kHz repetition rate. Until now, the waveform-controlled laser source delivered 1.2 mJ, 5 fs pulses with a relative CEP drift of 250 mrad RMS and a temporal contrast of $10^{-7}$ [7]. It was based on a Ti:Sa CPA system followed by post-compression in a gas-filled hollow-core fiber. When focused down to a spot size of 2 μm FWHM, the laser pulses could reach typical intensities on the solid target just below $10^{18}$ W/cm$^2$. Recent results obtained with this system include attosecond control of collective electron motion in plasmas, the generation of synchronized isolated attosecond pulses from plasma mirrors and the acceleration of energetic ions [8–10].

In this paper, we present the upgrade of the Salle Noire laser system, to pass into the relativistic intensity regime, i.e. with intensities on target well above $10^{18}$ W/cm$^2$ at 800 nm wavelength. It means increasing both the pulse energy and the temporal contrast by several orders of magnitude, while maintaining high repetition rate and full waveform control. Two major techniques are now widely used to decrease the intensity of the amplified spontaneous emission (ASE) background of existing CPA systems. Plasma mirrors can efficiently clean the pulse at the output of the laser chain at the expense of the available energy [11, 12]. Another option is to insert a temporal filter inside the laser chain, subsequent re-amplification enabling to recover the energy losses. In this case, efficient temporal cleaning is provided by a nonlinear effect, which calls for a double-CPA architecture [13]. At LOA, we have developed and qualified a nonlinear filter relying on cross-polarized wave (XPW) generation [14]. XPW generation is an achromatic and degenerated four wave mixing process relying on the anisotropy of the $\chi^3$ tensor in isotropic crystals. A linearly polarized input beam is focused onto a nonlinear crystal placed between crossed polarizers and generates an orthogonally polarized wave (XPW). The cubic dependence between output and input intensities provides an improvement in temporal contrast that is limited only by the extinction ratio of the crossed polarizers (3-4 orders of magnitude). Nowadays, XPW filter has become widespread in high peak-power laser systems [12, 15–17].

Therefore, we modified the laser system to a double-CPA architecture with an XPW filter. Another challenge was associated with this upgrade. As the final goal is to post-compress the laser pulses down to the few-cycle regime, CEP stability is needed. Consequently, the increase of the output energy requires specific dispersion management in the second CPA to reduce B-integral during amplification while preserving the CEP stability.

As a result, we demonstrate the realization of the first double-CPA CEP-stable laser system featuring 8 mJ, 22fs, 1 kHz pulses with a temporal contrast of 10$^{-11}$. The overall layout of the laser system appears in Fig. 1.

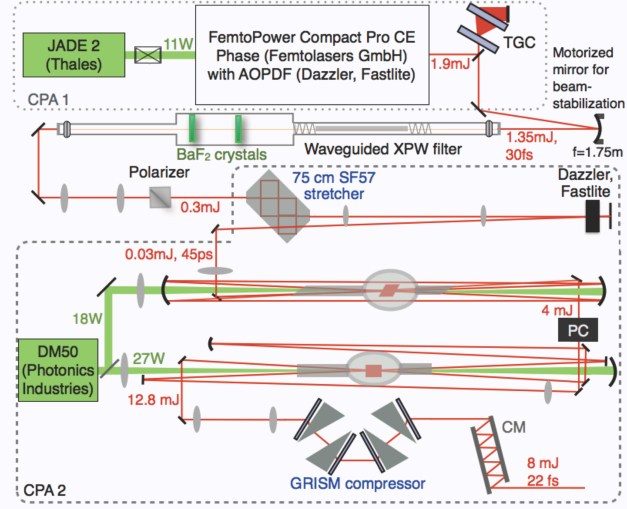

Fig. 1 Experimental scheme of high-contrast, 8 mJ, 22 fs, 1kHz CEP-stable double-CPA system. (TGC: transmission gratings compressor, PC: Pockels cell, CM: chirped mirrors).

Detailed description and characterization of the high-contrast injector can be found in [18]. It consists of a commercial 1 kHz CEP-stabilized CPA (CPA1, Femtopower Compact Pro CE Phase, Femtolasers GmbH) followed by an XPW contrast filter. The oscillator pulses (Rainbow) are CEP locked via pump-laser amplitude modulation and stretched in a 20 cm-long SF57 block before amplification up to 1.9 mJ in a ten-pass Ti:Sa amplifier including an AOPDF (low-jitter Dazzler HR800, Fastlite). The amplification stage is pumped by 11 W from a 1 kHz frequency-doubled Q-switched Nd:YLF laser (JADE, Thales). The 7 ps long amplified pulses are then compressed through a transmission-grating compressor (TGC) with an overall efficiency of 85%. The TGC consists of a pair of 1280 l/mm gratings used at the Littrow incidence angle (30.8°) and separated by 10 mm. The compressed pulse duration is 28 fs.

The pulses are then sent to the XPW filter which is an optimized version of the energy-scalable device described in [19]. It consists of a 47 cm, 250 μm inner diameter hollow-core waveguide providing efficient spatial filtering (75% transmission), followed by two thin, 1.5 mm thick, BaF$_2$ crystals with holographic crystallographic orientation, all under vacuum. The beam is then expanded to go through the output Glan polarizer enabling XPW pulse selection, with an extinction ratio of 3-4 orders of magnitude. 300 μJ pulses are routinely generated, corresponding to an XPW internal efficiency of 33% (fiber transmission and reflection losses on crystals deduced) and a global energy transmission of 22%. At this stage, the spectral bandwidth supports sub-10fs duration (Fig. 2(a)), the pulse presents a good spatio-temporal quality, the energy stability is around 2% RMS over several hours and the CEP drift can be kept below 200 mrad RMS [18].

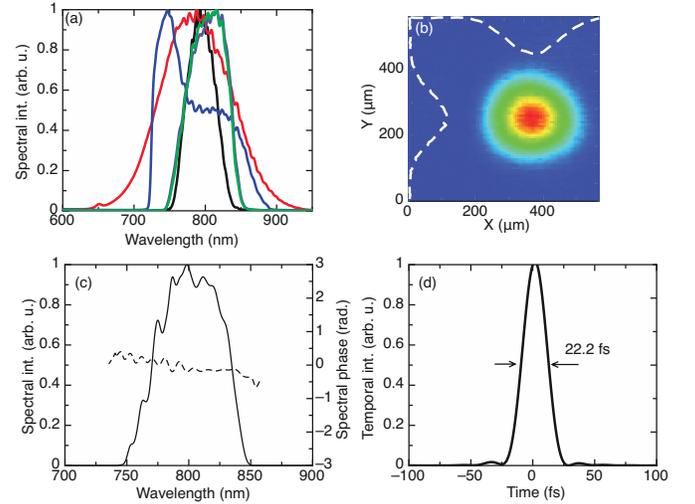

Fig. 2 (a) Evolution of the spectrum along the laser chain: after the FemtoPower (black), after XPW (red), after stretcher (bulk and Dazzler) of CPA2 (blue), after power amplifier (purple) and after the compressor (green). (b)Far-field spatial quality of the compressed pulse. Temporal measurement of the compressed 8mJ pulse (Wizzler) : (c) spectral amplitude and phase, (d) temporal intensity profile.

For the second CPA (CPA2), the issue of dispersion management must be carefully addressed. So far, in order to minimize CEP drift, compact stretcher (e.g. bulk) and compressor are preferred [20, 21]. The compressor has to present minimum optical beam path while an AOPDF can help to compensate for the spectral phase of the stretcher and amplifier material. The stretching factor should give a moderate B-integral value during amplification and nonlinear effects in the compressor should be avoided. The solution applied in CPA1 consists in a bulk stretcher and a TGC, whose chromatic dispersions do not match (TOD/GDD ratio is of opposite sign). Consequently, the finite pulse shaping capability of the AOPDF limits the stretching factor to about 0.1 ps/nm, thus preventing efficient amplification beyond a few mJ. Recently, we proposed an efficient and compact grism compressor design (combination of prisms and TG) able to match the first orders of the chromatic dispersion introduced by the bulk stretcher and amplifier and therefore relaxing the constraints on the AOPDF [22]. CEP stability was also demonstrated. The use of transmission gratings eases the geometrical constraints and the position of the prism pair between them can be used to tune the TOD/GDD ratio [23]. An air space between the prism and TG avoids final wavelength recombination inside the bulk and thus limits nonlinear effects. Here, we present a grism design optimized for high-energy pulses and larger stretching factor. An unfolded configuration is implemented to use smaller gratings, with a total footprint of 40*27 cm$^2$. The arrangement is shown in Fig. 3(a). It consists of 35*30 mm$^2$ and 70*30 mm$^2$, 1282 l/mm TG used at Littrow incidence, coupled to AR-coated, SF57 prisms (70*50 mm$^2$, apex 43.5°). 78% transmission is measured over 750 nm-850 nm spectral bandwidth. The introduced group delay was measured by inserting the grism in the arm of a Michelson interferometer (Fig. 3(b)). The stretching factor is 0.75 ps/nm, 7 times larger than in CPA1, with a TOD/GDD ratio of 0.3 fs. The chromatic dispersion can be decomposed as follows : -260,000 fs$^2$, -80,000 fs$^3$, -1.6e6 fs$^4$. This huge amount of FOD has to be compensated by an AOPDF and is the limit of the current design.

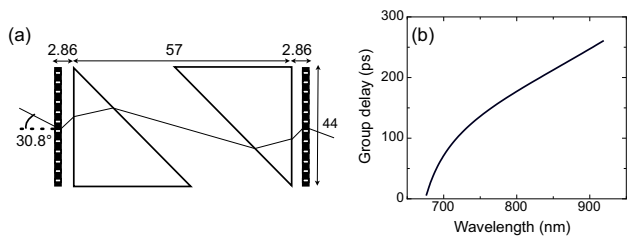

Fig. 3 (a) Grism design (distances are in mm). (b)Measured group delay introduced by two identical pairs of grism in an unfolded version.

The amplifier stages were first bypassed to check the validity of the compression without any amplification. The XPW beam is sent to the stretcher, a 75 cm-long SF57 bulk with a snooker design (117.8 * 151.5 * 25 mm$^3$, 6 total internal reflections) with 70% transmission. XPW beam diameter (20 mm) is sufficient to avoid any nonlinearity. The bulk is followed by a double-pass AOPDF (low-jitter Dazzler HR45, Fastlite). Although the Dazzler is able to transmit the full XPW spectral bandwidth, the FOD amount needed for compression imposes significant spectral clipping, as can be seen in Fig. 2(a). The overall transmission (bulk + Dazzler) is then 10% and the stretched pulse duration is about 45 ps FWHM. The pulse was then compressed in the grism to its Fourier limit of 18 fs.

Stretched pulse amplification in CPA2 is achieved in two stages. Both amplifiers are pumped by a 50 W, 1 kHz frequency-doubled Q-switched Nd:YLF laser (DM50, Photonics Industries) split into two arms and are cryogenically cooled under vacuum to 180 K. The 30 μJ seed pulses are first boosted up to 4 mJ after 6-passes amplification in a 8 mm Ti:Sa crystal pumped by 18 W. This booster stage, provided by Femtolasers GmbH, is a confocal multipass configuration. At this stage, gain narrowing reduces the spectral bandwidth (Fig. 2(a)). The final power amplifier consists in a home-made, image-relayed 2-passes in a 3 mm Ti:Sa crystal, cut at Brewster angle for S polarization. With 27 W pump power and both seed and pump beam diameter at 650 μm (1/e$^2$), the seed pulses are typically amplified to 9 mJ and 12.8 mJ respectively after each pass. A third pass can provide higher energy (14 mJ) but at the expense of the spatial beam quality (spatial modulations inherited from the pump beam profile) and this last pass is not implemented yet. Anyway, gain saturation is nearly reached with the second pass and energy stability is then 1% RMS over hours.

For compression in the grism, the beam size is then expanded to 22 mm (1/e$^2$), limited by the gratings size. To avoid nonlinearity in the last grating substrate, final compression is achieved by 10 bounces on highly dispersive chirped mirrors (-2000 fs$^2$). Therefore, no spectral modifications occur during compression (Fig. 2(a)). The overall transmission of the compressor is then 67%. Compressed pulse energy is 8 mJ. Self-referenced spectral interferometry (Wizzler, Fastlite) enables accurate spectral phase compensation via feedback loop on the Dazzler. We measure a pulse duration of 22 fs (Fig. 2(d)) (FT limited).

Far-field spatial quality is shown in Fig. 2(b) and a beam Strehl ratio better than 0.7 is measured (limited by astigmatism).

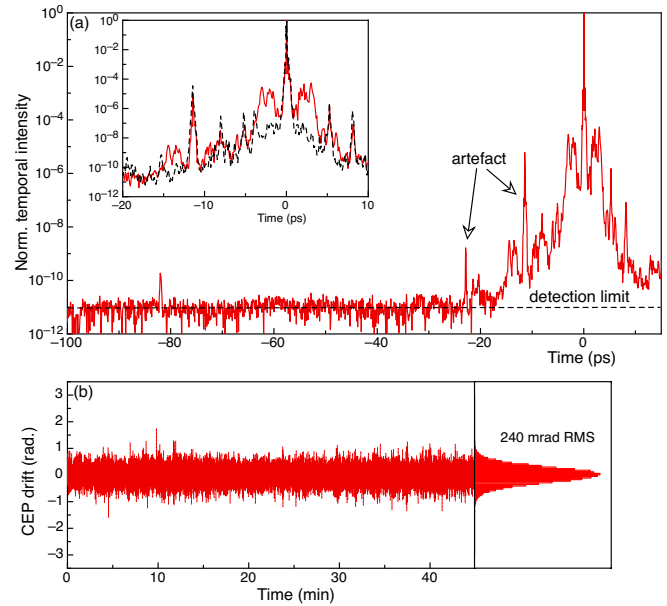

Fig. 4 (a) Measured temporal contrast of the 8 mJ, 22 fs pulses. The inset presents a zoom of the -20 ps / 10 ps temporal window. The dashed black line is the temporal contrast measured when applying a spectral Gaussian apodization. (b) Relative CEP drift with feedback control on the oscillator.

The temporal contrast of the 8 mJ, 22 fs pulses is measured thanks to a home-made high-dynamic range third-order autocorrelator. The energy is reduced to 2 mJ before entering the device. The incoherent contrast between -100 ps and -15 ps is 10$^{-11}$, limited by the detection (Fig. 4(a)). The improvement of 3 orders of magnitude compared to the Femtopower laser corresponds to the extinction ratio of the output polarizer of the XPW setup. One can notice that the main pulse is surrounded by a pedestal culminating at $10^{-5}$ relative intensity. We found that these coherent structures are due to a severe clipping of the spectral bandwidth in the Dazzler of CPA2. When applying a Gaussian apodization of the Dazzler acoustic wave to smooth the edges of the spectrum, at the expense of the final pulse duration (28 fs instead of 22 fs), the symmetric shoulders (± 2.5 ps) are decreased to $10^{-8}$ (inset Fig. 4(a)). Same contrast measurements have been obtained using a commercial device (Sequoia, Amplitude Technologies).

CEP stability of the whole system is the last critical feature to be characterized. To implement it, a few microjoules of the compressed pulse are split off and sent into a collinear f-to-2f interferometer (APS800, Menlo Systems). The slow CEP drift of the output pulses at 1kHz is pre-compensated by a feedback loop to the oscillator locking electronics. Results are shown on Fig. 4(b). The RMS phase error is 240 mrad over 45 minutes. This behavior is not affected by the apodization of the spectrum, nor by the energy level sent into the compressor.

To conclude, we have demonstrated a prototype laser source for driving relativistic optics at 1 kHz. We have shown CEP-stabilization of a double-CPA laser system and demonstrated a compact grism-based compression design suited to the compression of multi-mJ energy pulses. In a next step, post-compression in hollow-core fiber will be implemented to produce TW peak-power (4 mJ, 4 fs) controlled few-cycle light waveforms [24]. In this post-compression stage, CEP control is expected to be preserved, if specific care is devoted to keep a good spectral and energy stability.

Financial support from the Conseil Général de l'Essonne (ASTRE 2010 program), the Agence Nationale pour la Recherche (grants ANR-09-JCJC-0063 and ANR-11-EQPX-005-ATTOLAB) and Triangle de la Physique (RTRA, SOURCELAB-2012-057T) is gratefully acknowledged.

### References

[1] E. Esarey, C.B. Schroeder and W.P. Leemans, Rev. Mod. Phys. **81**, 1229 (2009).
[2] C. Thaury, F. Quéré, J.P. Geindre, A. Levy, T. Ceccotti, P. Monot, M. Bougeard, F. Rau, P. D'Oliveira, P. Audebert, R. Marjoribanks, P. Martin, Nat. Phys. **3**, 424 (2007).
[3] K. Ta Phuoc, S. Corde, C. Thaury, V. Malka, A. Tafzi, J.P. Goddet, R.C. Shah, S. Sebban and A. Rousse, Nat. Phot. **6**, 308 (2012).
[4] F. Krausz and M. Ivanov, Rev. Mod. Phys. **81**, 163–234 (2009).
[5] A. Buck, M. Nicolai, K. Schmid, C.M.S. Sears, A. Savert, J.M. Mikhailova, F. Krausz, M.C. Kaulza and L. Veisz, Nat. Phys. **7**, 548 (2011).
[6] P. Heissler, R. Hörlein, J.M. Mikhailova, L. Waldecker, P. Tzallas, A. Buck, K. Schmid, C.M.S. Sears, F. Krausz, L. Veisz, M. Zepf, and G.D. Tsakiris Phys. Rev. Lett. **108**, 235003 (2012).
[7] X. Chen, A. Malvache, A. Ricci, A. Jullien, and R. Lopez-Martens, Laser Physics **21**, 1–4 (2011).
[8] A. Borot, A. Malvache, X. Chen, A. Jullien, J.-P. Geindre, P. Audebert, G. Mourou, F. Quéré, and R. Lopez-Martens, Nat. Phys. **8**, 416 (2012).
[9] J. A. Wheeler, A. Borot, S. Monchoce, H. Vincenti, A. Ricci, A. Malvache, R. Lopez-Martens and F. Quéré, Nat. Phot. **6**, 829 (2012).
[10] M. Veltcheva, A. Borot, C. Thaury, A. Malvache, E. Lefebvre, A. Flacco, R. Lopez-Martens, and V. Malka, Phys. Rev. Lett. **108**, 075004 (2012).
[11] G. Doumy, F. Quéré, O. Gobert, M. Perdrix, P. Martin, P. Audebert, J. C. Gauthier, J.-P. Geindre, and T. Wittmann, Phys. Rev. E **69**, 026402 (2004).
[12] J. M. Mikhailova, A. Buck, A. Borot, K. Schmid, C. Sears, G. D. Tsakiris, F. Krausz, and L. Veisz, Opt. Lett. **36**, 3145–3147 (2011).
[13] M. P. Kalashnikov, E. Risse, H. Schönagel, and W. Sandner, Opt. Lett. **30**, 923–925 (2005).
[14] A. Jullien, O. Albert, F. Burgy, G. Hamoniaux, J.- P. Rousseau, J.-P. Chambaret, F. Augé-Rochereau, G. Chériaux, J. Etchepare, N. Minkovski, and S. M. Saltiel, Opt. Lett. **30**, 920–922 (2005).
[15] A. Flacco, F. Sylla, M. Veltcheva, M. Carrié, R. Nuter, E. Lefebvre, D. Batani, and V. Malka, Phys. Rev. E **81**, 036405 (2010).
[16] W. P. Leemans and C. Simon-Boisson, Conference of the International Committee on Ultra-High Intensity Lasers (ICUIL, 2012).
[17] G. Matras, F. Lureau, S. Laux, O. Casagrande, C. Radier, O. Chalus, F. Caradec, L. Boudjemaa, C. Simon-Boisson, R. Dabu, F. Jipa, L. Neagu, I. Dancus, D. Sporea, C. Fenic, C. Grigoriu, Advanced Solid State Lasers, 2013.
[18] A. Ricci, A. Jullien, J.P. Rousseau, R. Lopez-Martens, App. Sci. **3**, 314–324 (2013).
[19] A. Ricci, A. Jullien, J.P. Rousseau, Y. Liu, A. Houard, P. Ramirez, D. Papadopoulos, A. Pellegrina, P. Georges, F. Druon, N. Forget, R. Lopez-Martens, Rev. Sci. Inst. **84**, 043106 (2013).
[20] Z. Chang, Appl. Opt. **45**, 8350 (2006).
[21] L. Canova, X. Chen, A. Trisorio, A. Jullien, A. Assion, G. Tempea, N. Forget, T. Oksenhendler, and R. Lopez-Martens, Opt. Lett. **34**, 1333 (2009).
[22] A. Ricci, A. Jullien, N. Forget, V. Crozatier, P. Tournois, and R. Lopez-Martens, Opt. Lett. **37**, 1196–1198 (2012).
[23] N. Forget, V. Crozatier, P. Tournois, Appl. Phys. B 109, **121** (2012).
[24] F. Böhle, M. Kretschmar, A. Jullien, M. Kovacs, M. Miranda, R. Romero, H. Crespo, P. Simon, R. Lopez-Martens and T. Nagy, submitted (2014).

## References

[1] E. Esarey, C.B. Schroeder and W.P. Leemans, *Physics of laser-driven plasma-based electron accelerators*, Rev. Mod. Phys. **81**, 1229 (2009).

[2] C. Thaury, F. Quéré, J.P. Geindre, A. Levy, T. Ceccotti, P. Monot, M. Bougeard, F. Rau, P. D'Oliveira, P. Audebert, R. Marjoribanks, P. Martin, *Plasma mirrors for ultrahigh-intensity optics*, Nat. Phys. **3**, 424 (2007).

[3] K. Ta Phuoc, S. Corde, C. Thaury, V. Malka, A. Tafzi, J.P. Goddet, R.C. Shah, S. Sebban and A. Rousse, *All-optical Compton gamma-ray source*, Nat. Phot. **6**, 308 (2012).

[4] F. Krausz and M. Ivanov, *Attosecond physics*, Rev. Mod. Phys. **81**, 163–234 (2009).

[5] A. Buck, M. Nicolai, K. Schmid, C.M.S. Sears, A. Savert, J.M. Mikhailova, F. Krausz, M.C. Kaulza and L. Veisz, *Real-time observation of laser-driven electron acceleration*, Nat. Phys. **7**, 548 (2011).

[6] P. Heissler, R. Hörlein, J.M. Mikhailova, L. Waldecker, P. Tzallas, A. Buck, K. Schmid, C.M.S. Sears, F. Krausz, L. Veisz, M. Zepf, and G.D. Tsakiris, *Few-Cycle Driven Relativistically Oscillating Plasma Mirrors: A Source of Intense Isolated Attosecond Pulses*, Phys. Rev. Lett. **108**, 235003 (2012).

[7] X. Chen, A. Malvache, A. Ricci, A. Jullien, and R. Lopez-Martens, *Efficient hollow fiber compression scheme for generating multi-mJ, carrier-envelope phase stable, sub-5fs pulses,* Laser Physics **21**, 1–4 (2011).

[8] A. Borot, A. Malvache, X. Chen, A. Jullien, J.-P. Geindre, P. Audebert, G. Mourou, F. Quéré, and R. Lopez-Martens, *Attosecond control of collective electron motion in plasmas*, Nat. Phys. **8**, 416 (2012).

[9] J. A. Wheeler, A. Borot, S. Monchoce, H. Vincenti, A. Ricci, A. Malvache, R. Lopez-Martens and F. Quéré, *Attosecond lighthouses from plasma mirrors,* Nat. Phot. 6, 829 (2012).

[10] M. Veltcheva, A. Borot, C. Thaury, A. Malvache, E. Lefebvre, A. Flacco, R. Lopez-Martens, and V. Malka, *Brunel-dominated proton acceleration with a few-cycle laser pulse*, Phys. Rev. Lett. **108**, 075004 (2012).

[11] G. Doumy, F. Quéré, O. Gobert, M. Perdrix, P. Martin, P. Audebert, J. C. Gauthier, J.-P. Geindre, and T. Wittmann, *Complete characterization of a plasma mirror for the production of high-contrast ultraintense laser pulse*, Phys. Rev. E **69**, 026402 (2004).

[12] J. M. Mikhailova, A. Buck, A. Borot, K. Schmid, C. Sears, G. D. Tsakiris, F. Krausz, and L. Veisz, *Ultra-high-contrast few-cycle pulses for multipetawatt-class laser technology,* Opt. Lett. **36**, 3145–3147 (2011).

[13] M. P. Kalashnikov, E. Risse, H. Schönnagel, and W. Sandner, *Double chirped-pulse-amplification laser: a way to clean pulses temporally*, Opt. Lett. **30**, 923–925 (2005).

[14] A. Jullien, O. Albert, F. Burgy, G. Hamoniaux, J.-P. Rousseau, J.-P. Chambaret, F. Augé-Rochereau, G. Chériaux, J. Etchepare, N. Minkovski, and S. M. Saltiel, *$10^{-10}$ temporal contrast for femtosecond ultraintense lasers by cross-polarized wave generation*, Opt. Lett. **30**, 920–922 (2005).

[15] A. Flacco, F. Sylla, M. Veltcheva, M. Carrié, R. Nuter, E. Lefebvre, D. Batani, and V. Malka, *Dependence on pulse duration and foil thickness in high-contrast-laser proton acceleration*, Phys. Rev. E **81**, 036405 (2010).

[16] W. P. Leemans and C. Simon-Boisson, *The BELLA system and facility*, Conference of the International Committee on Ultra-High Intensity Lasers (ICUIL, 2012).

[17] G. Matras, F. Lureau, S. Laux, O. Casagrande, C. Radier, O. Chalus, F. Caradec, L. Boudjemaa, C. Simon-Boisson, R. Dabu, F. Jipa, L. Neagu, I. Dancus, D. Sporea, C. Fenic, C. Grigoriu, *First sub-25fs PetaWatt laser system*, Advanced Solid State Lasers, 2013.

[18] A. Ricci, A. Jullien, J.P. Rousseau, R. Lopez-Martens, *Front-end light source for a waveform-controlled high-contrast few-cycle laser system for high-repetition rate relativistic optics*, App. Sci. **3**, 314–324 (2013).

[19] A. Ricci, A. Jullien, J.P. Rousseau, Y. Liu, A. Houard, P. Ramirez, D. Papadopoulos, A. Pellegrina, P. Georges, F. Druon, N. Forget, R. Lopez-Martens, *Energy-scalable temporal cleaning device for femtosecond laser pulses based on cross-polarized wave generation*, Rev. Sci. Inst. **84**, 043106 (2013).

[20] Z. Chang, *Carrier-envelope phase shift caused by grating-based stretchers and compressors*, Appl. Opt. **45**, 8350 (2006).

[21] L. Canova, X. Chen, A. Trisorio, A. Jullien, A. Assion, G. Tempea, N. Forget, T. Oksenhendler, and R. Lopez-Martens, *Carrier-envelope phase stabilization and control using a transmission grating compressor and an AOPDF,* Opt. Lett. **34**, 1333 (2009).

[22] A. Ricci, A. Jullien, N. Forget, V. Crozatier, P. Tournois, and R. Lopez-Martens, *Grism compressor for carrier-envelope phase-stable millijoule-energy chirped pulse amplifier lasers featuring bulk material stretcher*, Opt. Lett. **37**, 1196–1198 (2012).

[23] N. Forget, V. Crozatier, P. Tournois, *Transmission Bragg-grating grisms for pulse compression*, Appl. Phys. B **109**, 121 (2012).

[24] F. Böhle, M. Kretschmar, A. Jullien, M. Kovacs, M. Miranda, R. Romero, H. Crespo, P. Simon, R. Lopez-Martens and T. Nagy, *Compression of CEP-stable multi-mJ laser pulses down to 4fs in long hollow fibers*, submitted (2014).